# *DcardNet*: Diabetic Retinopathy Classification at Multiple Levels Based on Structural and Angiographic Optical Coherence Tomography

Pengxiao Zang, Liqin Gao, Tristan T. Hormel, Jie Wang, Qisheng You, Thomas S. Hwang, and Yali Jia*

*Abstract— Objective:* Optical coherence tomography (OCT) and its angiography (OCTA) have several advantages for the early detection and diagnosis of diabetic retinopathy (DR). However, automated, complete DR classification frameworks based on both OCT and OCTA data have not been proposed. In this study, a convolutional neural network (CNN) based method is proposed to fulfill a DR classification framework using *en face* OCT and OCTA. *Methods:* A densely and continuously connected neural network with adaptive rate dropout (*DcardNet*) is designed for the DR classification. In addition, adaptive label smoothing was proposed and used to suppress overfitting. Three separate classification levels are generated for each case based on the International Clinical Diabetic Retinopathy scale. At the highest level the network classifies scans as referable or non-referable for DR. The second level classifies the eye as non-DR, non-proliferative DR (NPDR), or proliferative DR (PDR). The last level classifies the case as no DR, mild and moderate NPDR, severe NPDR, and PDR. *Results:* We used 10-fold cross-validation with 10% of the data to assess the network's performance. The overall classification accuracies of the three levels were 95.7%, 85.0%, and 71.0% respectively. *Conclusion/Significance:* A reliable, sensitive and specific automated classification framework for referral to an ophthalmologist can be a key technology for reducing vision loss related to DR.

## I. INTRODUCTION

OPTICAL coherence tomography (OCT) can generate depth-resolved, micrometer-scale-resolution images of ocular fundus tissue based on reflectance signals obtained using interferometric analysis of low coherence light [1]. By scanning multiple B-frames at the same position, change in the OCT reflectance properties can be measured as, e.g., decorrelation values to differentiate vasculature from static tissues. This technique is called OCT angiography (OCTA), and it can provide high-resolution images of the microvasculature of retina [2, 3]. Numerous investigators explored OCTA in the detection and diagnosis of various ocular diseases, and demonstrated many advantages when compared to traditional imaging modalities such as fundus photography or fluorescein angiography [3]. Among these is diabetic retinopathy (DR), which affects the retinal capillaries and is a leading cause of preventable blindness globally [4]. OCT-based biomarkers such as central macular thickness and OCTA-based biomarkers such as avascular areas have demonstrated superior potential for diagnosing and classifying DR compared to traditional imaging modalities [5-8]. However, recently emerged automated deep-learning classification methods were largely based on color fundus photography (CFP) [9-12]. Therefore, taking advantages of both powerful deep learning tools and innovative structural and angiographic information, we developed an automated framework that can perform a full DR classification (across datasets including all DR grades) based on *en face* OCT and OCTA projected from the same volumetric scans.

In order to improve classification accuracy and reliability, a new convolutional neural network architecture was designed based on dense and continuous connection with adaptive rate dropout (*DcardNet*). The system produces three classification levels to fulfill requests in clinical diagnosis. Non-referable and referable DR (nrDR and rDR) are classified in the first level. No DR, non-proliferative DR (NPDR), and proliferative DR (PDR) are in the second classification level. No DR, mild and moderate NPDR, severe NPDR, and PDR are in the third level. While training *DcardNet*, adaptive label smoothing was used to reduce overfitting. To improve interpretability and help understand which regions contribute to the diagnosis, class activation maps (CAM) were also generated for each DR class [13].

## II. RELATED WORKS

Several methods for the automated classification of DR severity have been proposed since the convolutional neural network (CNN) became the most widely used solution for image classification problems [9-12, 14-17]. Most of these

This work was supported by grant R01 EY027833, R01 EY024544, P30 EY010572 from the National Institutes of Health (Bethesda, MD), and an unrestricted departmental funding grant and William & Mary Greve Special Scholar Award from Research to Prevent Blindness (New York, NY).

P. Zang, J. Wang, and *Y. Jia are with Casey Eye Institute, Oregon Health & Science University, Portland, OR, USA, and also with Department of Biomedical Engineering, Oregon Health & Science University, Portland, OR, USA (correspondence e-mail: jiaya@ohsu.edu).
T. T. Hormel, Q You, and T. S. Hwang are with Casey Eye Institute, Oregon Health & Science University, Portland, OR, USA.
L. Gao is with Casey Eye Institute, Oregon Health & Science University, Portland, OR, USA, and also with Beijing Tongren Eye Center, Beijing Key Laboratory of Ophthalmology and Visual Science, Beijing Tongren Hospital, Capital Medical University, Beijing, China.

methods are based on CFP, which is a traditional and commonly used technique capable of DR diagnosis. R. Gargeya *et al*. proposed a machine learning based method to classify CFP images as healthy (no retinopathy) or having DR [9]. They used a customized *ResNet* architecture [18] to extract features from the input CFPs. The final classification was performed on a decision tree classification model by using the combination of extracted features and three metadata variables. They achieved a 0.97 area under receiver operating curve (AUC) after 5-fold stratified cross-validation. In addition, a visualization heatmap was generated for each input CFP based on visualization layer in the end of their network [13]. V. Gulshan *et al*. used Inception-v3-based transfer learning to classify the CFP mainly as rDR and nrDR [11]. In the validation tests on two publicly available datasets (eyePACS-1 and Messidor-2), they achieved an AUC of 0.991 and 0.990, respectively. M. D. Abramoff *et al*. also proposed a CNN-based method to classify CFP images as rDR and nrDR and achieved an AUC of 0.980 during validation [10]. For more detailed DR classification, R. Ghosh *et al.* proposed a CNN-based method to classify the CFP images into both two-class (no DR vs DR) and five severities: no DR, mild NPDR, moderate NPDR, severe-NPDR, and PDR [12]. They achieved an overall accuracy of 85% for the classification into five severities.

However, all of the above methods were based on the CFP. Compared to CFP, OCT and OCTA can provide more detailed information (i.e. 3D, high-resolution, vascular and structural imaging). An automated DR classification framework based on OCT/OCTA could reduce the number of procedures that must be performed in the clinic if OCT/OCTA can deliver the same diagnostic value as other modalities, which will ultimately reduce clinical burden and healthcare costs. Therefore, an automated framework for DR classification based on OCT and OCTA data is desirable.

H. S. Sandhu *et al*. proposed a computer-assisted diagnostic (CAD) system based on quantifying three OCT features: retinal reflectivity, curvature, and thickness [14]. A deep neural network was used to classify each case as no DR or NPDR based on those three retinal features and achieved an overall accuracy of 93.8%. The same group also proposed a CAD system for DR classification based on quantified features from OCTA [15]: blood vessel density, foveal avascular zone (FAZ) area, and blood vessel caliber and trained a support vector machine (SVM) with a radial basis function (RBF) kernel. They achieved an overall accuracy of 94.3%. However, these systems examined and classified only no DR and NPDR cases. M. Alam *et al*. proposed a support vector machine-based DR classification CAD system using six quantitative features generated from OCTA: blood vessel tortuosity, blood vascular caliber, vessel perimeter index, blood vessel density, foveal avascular zone area, and foveal avascular zone contour irregularity [16]. They achieved 94.41% and 92.96% accuracies for control versus disease (NPDR) and control versus mild NPDR. In addition, they achieved 83.94% accuracy for multiclass classification (control, mild NPDR, moderate NPDR, and severe NPDR). However, as only pre-determined features were incorporated into this model, it could not learn from the much richer feature space latent in the entire OCTA data. In addition, CAD systems based on only empirically selected biomarkers have limited potential for further improvements even as the number of available datasets grows. M. Heisler *et al.* proposed a DR classification method based on *en face* OCT and OCTA images using ensemble networks [17]. Each case was classified as nrDR or rDR and they achieved an overall accuracy of 92.0%. In addition, the CAM of each *en face* image was generated. However, only 2-class classification was performed in this study. Therefore, an OCT and OCTA based DR classification framework capable of fulfilling different clinical requests and generating CAMs is needed.

There are two major challenges for OCT and OCTA-based DR classification. First, OCTA generates a much greater detailed image of the vasculature than traditional CFP. Extracting classification related features from such detailed information is much more challenging compared with the CFP-based classification. The second challenge is the relatively small size of the available OCT and OCTA dataset, compared to the very large CFP dataset used in the previous CFP-based networks. This challenge can lead to a severe overfitting problem during the training of the network. Addressing these challenges requires a network architecture with not only efficient convergence but also low overfitting. We designed a densely and continuously connected neural network with adaptive rate dropout and used it to perform a DR classification in three levels. We also produced corresponding CAMs in this study. In addition, adaptive label smoothing was proposed to further reduce overfitting. The main contributions of the present work are as follows:

- We present an automated framework for the DR classification and CAM generation based on both OCT and OCTA data. In this framework, three DR classification levels are performed for the first time.
- We propose a new network architecture based on dense and continuous connections with adaptive rate dropout.
- We propose an adaptive label smoothing to suppress overfitting and improve the performance generalization of the trained network.

III. MATERIALS

In this study, 303 eyes from 250 participants, including healthy volunteers and patients with diabetes (with or without DR) were recruited and examined at the Casey Eye Institute, Oregon Health & Science University. Masked trained retina specialists graded the disease severity based on Early Treatment of Diabetic Retinopathy Study (ETDRS) scale [19] using corresponding 7-field fundus photography. Based on the recent studies on referable retinopathy level shown in the International Clinical Diabetic Retinopathy scale [20], we defined referable retinopathy as the equivalent ETDRS grade, which is grade 35 or worse. The participants were enrolled after informed consent in accordance with an Institutional Review Board (IRB # 16932) approved protocol. The study was conducted in compliance with the Declaration of Helsinki and Health Insurance Portability and Accountability Act.

The macular region of each eye was scanned once or twice (after a one-year gap) using a commercial 70-kHz spectral-domain OCT (SD-OCT) system (Avanti RTVue-XR, Optovue Inc) with 840-nm central wavelength. The scan regions were 3.0 × 3.0 mm and 1.6 mm in depth (304 × 304 ×

640 pixels) centered on the fovea. Two repeated B-frames were captured at each line-scan location to calculate the OCTA decorrelation values. The blood flow of each line-scan location was detected using the split-spectrum amplitude-decorrelation angiography (SSADA) algorithm based on the speckle variation between two repeated B-frames [2, 21]. The OCT structural images were obtained by averaging two repeated B-frames. For each data set, two volumetric raster scans (one x-fast scan and one y-fast scan) were registered and merged through an orthogonal registration algorithm to reduce motion artifacts [22].

For each pair of OCT and OCTA data, the following retinal layers were automatically segmented (Fig. 1) based on the commercial software in the SD-OCT system (Avanti RTVue-XR, Optovue Inc): inner limiting membrane (ILM), nerve fiber layer (NFL), ganglion cell layer (GCL), inner plexiform layer (IPL), inner nuclear layer (INL), outer plexiform layer (OPL), outer nuclear layer (ONL), ellipsoid zone (EZ), retinal pigment epithelium (RPE), and Bruch's membrane (BM). In addition, for the cases with severe pathologies, the automated layer segmentation was manually corrected by graders using the customized COOL-ART software [23].

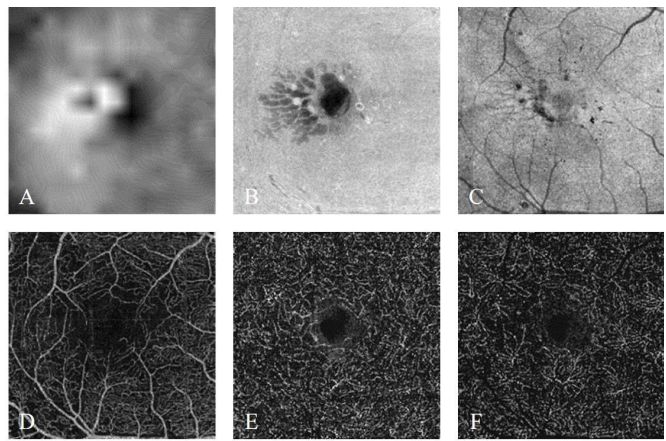

Fig. 2. The six input channels based on the OCT and OCTA data scanned from a moderate NPDR participant. (A) Inner retinal thickness map. (B) Inner retinal *en face* average projection. (C) Ellipsoid zone (EZ) *en face* average projection. (D) Superficial vascular complex (SVC) *en face* maximum projection. (E) Intermediate capillary plexus (ICP) *en face* maximum projection. (F) Deep capillary plexus (DCP) *en face* maximum projection.

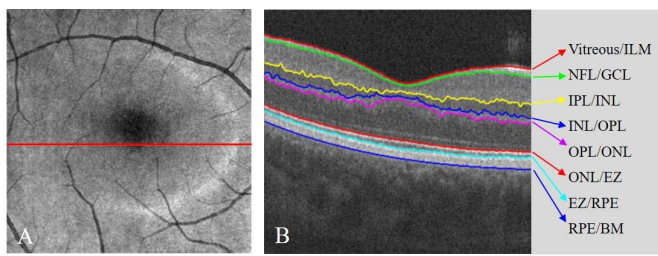

Fig. 1. The automated retinal layer segmentation from OCT structural image scanned from a healthy participant. (A) The *en face* average projection of the whole OCT structure. (B) The B-frame corresponding to the position of red line in (A). The eight boundaries of the seven main retinal layers were segmented.

Based on the segmented boundaries, six *en face* projections from OCT reflectance signals and OCTA decorrelation values were obtained and used to build a six-channel input data (Fig. 2). The first three channels were the inner retinal thickness map (z-axis distance between the Vitreous/ILM and OPL/ONL), inner retinal *en face* average projection (Vitreous/ILM to OPL/ONL) and EZ *en face* average projection (ONL/EZ to EZ/RPE) based on the volumetric OCT (Fig. 2A-C). The last three channels were the *en face* maximum projections of the superficial vascular complex (SVC), intermediate capillary plexus (ICP), and deep capillary plexus (DCP) based on the volumetric OCTA. (Fig. 2D-F) [24]. The SVC was defined as the inner 80% of the ganglion cell complex (GCC), which included all structures between the ILM and IPL/INL border. The ICP was defined as the outer 20% of the GCC and the inner 50% of the INL. The DCP was defined as the remaining slab internal to the outer boundary of the OPL [6, 25]. In addition, the projection-resolved (PR) OCTA algorithm was applied to all OCTA scans to remove flow projection artifacts in the deeper plexuses [26, 27].

Three classification levels of each input data were built based on the ETDRS grades as scored by three ophthalmologists (Fig. 3). The first label was for 2 classes: nrDR and rDR. The second label was for 3 classes: no DR, NPDR and PDR. The last label was for 4 classes: no DR, mild and moderate NPDR, severe NPDR and PDR. Mild and moderate NPDR were not separated due to a lack of measurements on eyes with NPDR from which to procure make a balanced dataset. For each level, follow up scans (scanned after a one-year gap) that did not have a class change were removed from the dataset for corresponding level to avoid correlation. Therefore, number of scans for each classification level was different (Table I).

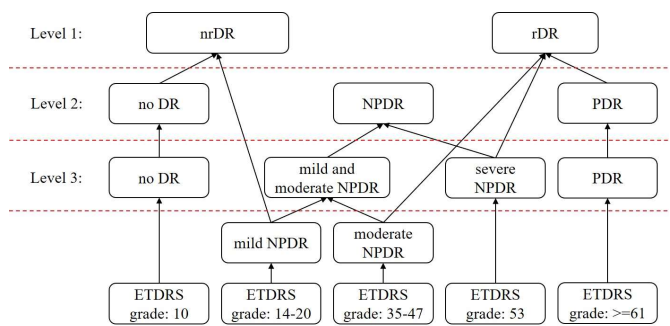

Fig. 3. The relations between the ETDRS grades and three levels of DR classifications.

TABLE I
DATA DISTRIBUTIONS OF THREE CLASSIFICATION LEVELS

| Classifications | Number of scans | Whole data size |
|---|---|---|
| nrDR | 95 | 294 |
| rDR | 199 | |
| no DR | 85 | 298 |
| NPDR | 128 | |
| PDR | 85 | |
| no DR | 85 | 302 |
| mild and moderate NPDR | 82 | |
| severe NPDR | 50 | |
| PDR | 85 | |

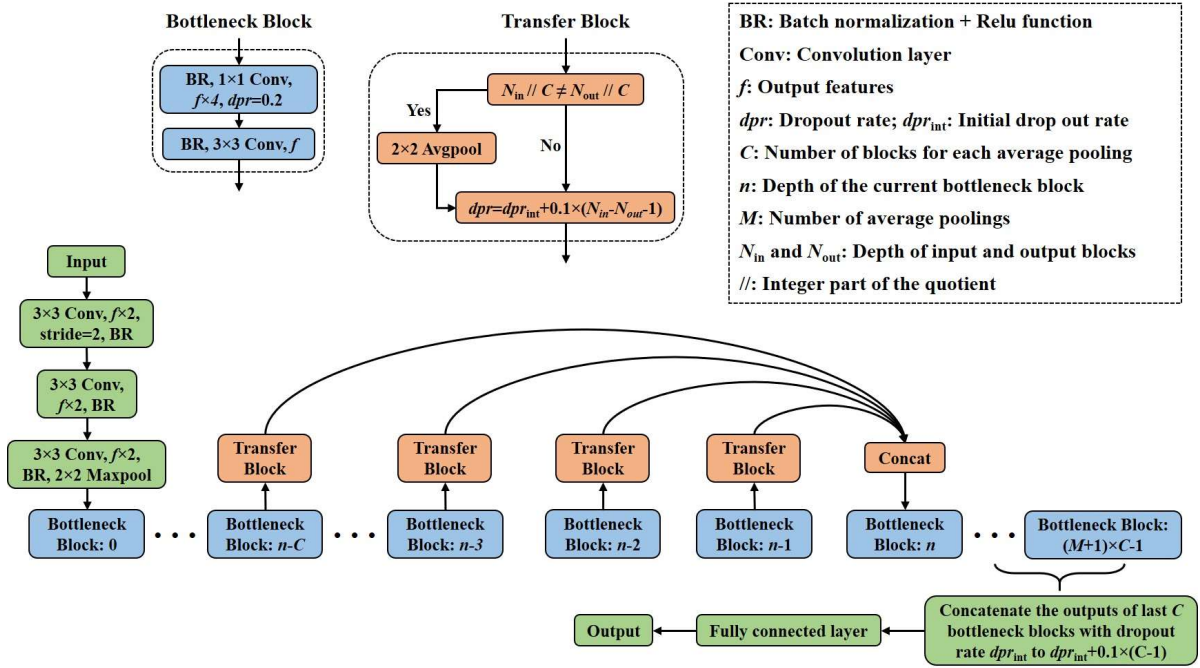

Fig. 4. The network architecture of the proposed *DcardNet*.

## IV. METHODS

The architecture of the *DcardNet* is shown in Fig. 4. The main feature of this architecture is that the input tensor for each bottleneck block was the concatenation of the output tensors from at most the *C* previous bottleneck blocks with adaptive dropout rates. The dropout rate [28] of each bottleneck was adaptively adjusted based on the distance between the depths of this block and the block to be calculated next. In addition, the size (height and width) of the output tensor was halved *M* times through transfer blocks to perform down-sampling. Detailed information for this method is described below.

### A. Bottleneck block

A 1×1 convolution is widely used as a bottleneck layer before 3×3 convolutions to improve the computational efficiency by reducing the number of input features [29]. Our network uses two convolutional layers in the bottleneck block. A 1×1 convolution layer with $f \times 4$ output features and 0.2 dropout rate [28] was used as the first convolutional layer. The second convolutional layer in the bottleneck block is a 3×3 convolution with *f* output features. In addition, a batch normalization [30] and *Relu* activation function [31, 32] were used before each convolutional layer.

### B. Transfer block

Before the concatenation of the output tensors from at most the last *C* bottleneck blocks, a transfer block was used to perform the adaptive rate dropout. The dropout rate (*dpr*) of the output tensor from each bottleneck block was calculated as

$$dpr = dpr_{int} + 0.1 \times (N_{in} - N_{out} - 1) \quad (1)$$

where $dpr_{int}$ is the initial dropout rate, $N_{out}$ is the depth of each bottleneck block which is to be concatenated, and $N_{in}$ is the depth of the bottleneck block that will use the concatenated tensor as input. In order to fulfill the down-sampling, the size of the tensor is halved before dropout using 2×2 average pooling if the integer part of the quotients between $N_{out}/C$ and $N_{in}/C$ were not equal.

### C. Dense and continuous connection with adaptive dropout

Dense connectivity has been proposed by G. Huang *et al.* [29] and used in *DenseNet* to improve information flow. However, the dense connection was only used within each dense block, not the whole network. In the *DcardNet*, the dense connection was continuously used in the whole network to further improve the information flow. In addition, the size and weight of each concatenated bottleneck block was adaptively adjusted using the transfer block to fulfill down-sampling and differentiate the importance of the information in different bottleneck blocks. The input tensor to each bottleneck block was

$$x_n^{in} = \text{concat}\left[T(x_{n-1}^{out}), T(x_{n-2}^{out}), \ldots, T(x_{\max(0, n-C)}^{out})\right] \quad (2)$$

where $x_n^{in}$ and $x_n^{out}$ are the input and output tensors of the $n_{th}$ bottleneck block, $\text{concat}[\bullet]$ is the concatenation operation, and $T(\bullet)$ is the transfer block.

### D. Adaptive label smoothing and data augmentation

The goal of training the network is high overall classification accuracy, defined as

$$Acc = \frac{1}{Num} \times \sum_{i=1}^{Num} a_i$$
$$a_i = \begin{cases} 1 & \arg\max(g_i) = \arg\max(p_i) \\ 0 & \text{otherwise} \end{cases} \quad (3)$$

where $g_i$ and $p_i$ are the $i^{th}$ ground truth and predicted labels at a given classification level, respectively, and $Num$ is the number of scans in the dataset. However, network parameters were optimized by minimizing the negative cross entropy loss

$$loss(g, p) = -\sum_{i=1}^{K} p_i \times \log g_i \quad (4)$$

where $K$ was the number of classes. According to (3), the prediction will always be right as long as the location of the largest value in the predicted label is the same as the ground truth label. Once this has been achieved, continuing to reduce the negative cross entropy loss only marginally improves the overall classification accuracy, and may lead to overfitting [33, 34]. Therefore, in this study, each ground truth label was gradually smoothed by an amount $s$ based on the class differences between the true class and false classes. Since class labels were sorted along a scale of DR severity, the smoothed class labels respect the decreasing likelihood that the label was misidentified. The labels at all three levels were smoothed according to

$$g_i = \begin{cases} 1.0 - s_i & \text{true class} \\ s_i \times \frac{\frac{1}{|t_j - t_i|}}{\sum_{j=1}^{K-1} \frac{1}{|t_j - t_i|}} & \text{other classes} \end{cases} \quad (5)$$

where $s_i$ is the reduction in the value of true class, and $t_j$ and $t_i$ respectively were the indexes of each incorrect class and the true class in $i^{th}$ label.

Variation between different OCTA data sets is intrinsically high. Some inputs converge well in a short time, but the convergence of other inputs might change significantly and repeatedly. According to the gradient of the weight variables in the network (6), the weights $w$ will converge to an input faster when the difference between the predication and corresponding ground truth label gets larger, and slower when the difference is smaller:

$$\frac{\partial loss}{\partial w} = \frac{1}{Num} \sum_{i=1}^{Num} x_i (p_i - g_i) \quad (6)$$

where $x_i$ is the $i^{th}$ input, $p_i$ and $g_i$ are the corresponding prediction and ground truth. In order to further increase the rate of convergence on the mispredicted inputs and decrease the rate of convergence on the correctly predicted inputs, the label smoothing value $s$ for each label was adaptively adjusted based on the prediction results during each training step according to

$$s_i = \begin{cases} \min(s_i + d, s_{max}) & \arg\max(g_i) = \arg\max(p_i) \\ \max(s_i - d, 0.0) & \text{otherwise} \end{cases} \quad (7)$$

where $s_i$ is the smoothing value for the $i^{th}$ label, and $d$ is an adjustment for each $s_i$ and $s_{max}$ was the upper limit of the smoothing value. Based on (7), the convergence rate of the inputs which were correctly predicted during each training iteration would be much lower than the other inputs.

In addition, no class weight balancing was used in training because adaptive label smoothing can achieve the same effect. Class weight balancing can tell the model to pay more attention to samples from an under-represented class by appropriately weighting the loss function to compensate for data deficiencies during training. Alternatively, the same effect could be achieved by smoothing the ground truth labels while maintaining the loss function (since classes with small label differences will contribute less to the loss). This is the approach taken in adaptive label smoothing, which has the additional advantage of allowing the smoothing function to updated during training to expedite balanced convergence.

Data augmentation is another method used for improving the performance generalization of a trained network. In this study, the number of training datasets was increased by a factor of 8 by including combinations of 90° rotations and horizontal and vertical flips (there is a grand total of 7 unique combinations of these transformations available). In order to make sure the selected inputs in each training batch were based on different cases, only one of the data augmented patterns (including the original inputs) was randomly chosen for each input during each training batch selection.

### E. Implementation details

The maximum number of the concatenated bottleneck blocks $C$ was set to 4. The number of output features $f$ after each bottleneck block was set to 24. $M$ was set to 3 which meant overall 16 bottleneck blocks were used in this architecture. This specific architecture is called *DcardNet*-36 which means overall 35 convolutional layers and 1 fully connected layer were used in the whole network, which yields 9264960 trainable parameters (Table II). In addition, for the 2-class, 3-class and 4-class DR classifications, the initial label smoothing value $s_i$ were set to 0.05, 0.005 and 0.005, adjusting steps $d$ were empirically chosen as 0.001, 0.0001 and 0.0001, and upper limits $s_{max}$ were set to 0.1, 0.01 and 0.01, respectively.

In order to ensure the credibility of the overall accuracy, 10-fold cross-validation was used on the DR classification at each level. In each fold, 10% of the data (with the same class distribution as the overall data set) was split on a patient-wise basis (scans from same patient only included in one set) and used exclusively for testing. The parameters were optimized by a stochastic gradient descent optimizer with Nesterov momentum (momentum = 0.9). During the training process, a batch size of 10 was empirically chosen and the total training steps for the three-level DR classification were set to 8000. In addition, an initial learning rate $lr_{init} = 0.01$ with cosine decay was used in this study [35]:

$$lr_{curr} = lr_{init} \times (0.97 \times d + 0.03)$$
$$d = \frac{1}{2}\left[1 + \cos(\pi \times step_{curr} / step_{stop})\right] \quad (8)$$

where $lr_{curr}$ was the current learning rate, $step_{curr}$ was the current training step and $step_{stop}$ was the step at which the learning rate ceased to decline. In this study, the $step_{stop}$ was empirically chosen as 6000.

TABLE II
ARCHITECTURE OF THE DCARDNET-36

| # | Input | Operator | Output |
|---|---|---|---|
| 1 | $224^2 \times 6$ | $3 \times 3$ conv2d, stride 2 | $112^2 \times 48$ |
| 2 | $112^2 \times 48$ | $3 \times 3$ conv2d, stride 1 | $112^2 \times 48$ |
| 3 | $112^2 \times 48$ | $3 \times 3$ conv2d, stride 1 | $112^2 \times 48$ |
| 4 | $112^2 \times 48$ | $3 \times 3$ max pool, stride 2 | $56^2 \times 48$ |
| b0 | $56^2 \times 48$ | Bottleneck block | $56^2 \times 24$ |
| b1 | t(b0), $56^2 \times 24$ | Bottleneck block | $56^2 \times 24$ |
| b2 | c[t(b0) - t(b1)], $56^2 \times 48$ | Bottleneck block | $56^2 \times 24$ |
| b3 | c[t(b0) - t(b2)], $56^2 \times 72$ | Bottleneck block | $56^2 \times 24$ |
| b4 | c[t(b0) - t(b3)], $28^2 \times 96$ | Bottleneck block | $28^2 \times 24$ |
| b5 | c[t(b1) - t(b4)], $28^2 \times 96$ | Bottleneck block | $28^2 \times 24$ |
| b6 | c[t(b2) - t(b5)], $28^2 \times 96$ | Bottleneck block | $28^2 \times 24$ |
| b7 | c[t(b3) - t(b6)], $28^2 \times 96$ | Bottleneck block | $28^2 \times 24$ |
| b8 | c[t(b4) - t(b7)], $14^2 \times 96$ | Bottleneck block | $14^2 \times 24$ |
| b9 | c[t(b5) - t(b8)], $14^2 \times 96$ | Bottleneck block | $14^2 \times 24$ |
| b10 | c[t(b6) - t(b9)], $14^2 \times 96$ | Bottleneck block | $14^2 \times 24$ |
| b11 | c[t(b7) - t(b10)], $14^2 \times 96$ | Bottleneck block | $14^2 \times 24$ |
| b12 | c[t(b8) - t(b11)], $7^2 \times 96$ | Bottleneck block | $7^2 \times 24$ |
| b13 | c[t(b9) - t(b12)], $7^2 \times 96$ | Bottleneck block | $7^2 \times 24$ |
| b14 | c[t(b10) - t(b13)], $7^2 \times 96$ | Bottleneck block | $7^2 \times 24$ |
| b15 | c[t(b11) - t(b14)], $7^2 \times 96$ | Bottleneck block | $7^2 \times 24$ |
| 21 | c[b12 – b15], $7^2 \times 96$ | Global average pool | 96 |
| 22 | 96 | Fully connected layer | 2/3/4 |

c[t(b0) - t(b3)] means concatenate c[] each output of bottleneck blocks b0 to b3 after transfer block t().

Both training and testing were implemented in Tensorflow version 1.13 on Windows 10 (64 Bit) platform. The workstation used in this study has an Intel (R) Core (TM) i7-8700K CPU @ 3.70GHz, 64.0 GB RAM and NVIDIA RTX 2080 GPU. The training time was 7 minutes for each training process (70 minutes for 10-fold cross-validation) and the inference time for a new case was 8 seconds.

## V. EXPERIMENTS

The overall prediction accuracy (the number of correctly predicted case divided by the number of whole data set) and corresponding 95% confidence interval (95% CI) varied across the three classification levels (Table III). In addition, the 10 models trained during the 10-fold cross validation were also used to predict on a balanced external dataset with 30 scans to further demonstrate the generalization of our DR classification framework. The overall accuracies of 2-class, 3-class, and 4-calss DR classification on the external dataset are 93.3% ± 2.4%, 82.7% ± 2.8%, and 68.7% ± 3.8%, respectively. Though the accuracies on the external dataset are about 2% - 3% lower than the accuracies on our local testing dataset, the results still

TABLE III
DR CLASSIFICATION ACCURACY AT MULTIPLE LEVELS

| | 2-class | 3-class | 4-class |
|---|---|---|---|
| 10-fold Accuracy (mean ± std) | 95.7% ± 3.9% | 85.0% ± 3.6% | 71.0% ± 4.8% |
| 95% CI | 93.3% - 98.1% | 82.8% - 87.2% | 68.0% - 74.0% |

show that our DR classification framework has a strong generalization on external dataset.

The sensitivity and specificity for each severity class in all three DR classification levels also varied and is shown in Table IV. The classification sensitivity of the severe NPDR was much lower than other classes. This is because the differences between adjacent levels of severity are much smaller than the variations between no DR, NPDR and PDR. In addition, the number of severe NPDR cases was also much smaller than other classes.

We also produced CAMs of inputs with different DR classes (Fig. 5), indicating the network's attention within the different DR classes. The macular regions with high positive values in the CAMs indicate they have high positive influences on the classification for the true class. On the contrary, the regions with nearly zero values in the CAMs have no or negative influence on the classification. In CAMs of cases without DR and cases with PDR regions close to the fovea had the highest positive influences on the classification. However, the vasculature around the fovea had the highest positive influences on the classification of NPDR cases. This difference may be caused by the appearance of features like fluids or non-perfusion areas. Overall, the areas with higher values (yellow to red) in the CAM were the regions the network used for decision making. By considering the CAMs, a doctor could judge the reasonableness of the automated DR classification and pay more attention on the high-value-areas during the diagnosis.

TABLE IV
SENSITIVITY AND SPECIFICITY OF EACH CLASS IN THREE DR CLASSIFICATION LEVELS

| Classification levels | DR severities | Sensitivities (mean, 95% CI) | Specificities (mean, 95% CI) |
|---|---|---|---|
| 2-class | nrDR | 91.0%, 86.4% - 95.6% | 98.0%, 96.4% - 99.6% |
| | rDR | 98.0%, 96.4% - 99.6% | 91.0%, 86.4% - 95.6% |
| 3-class | no DR | 86.7%, 81.3% - 92.1% | 93.3%, 91.8% - 94.8% |
| | NPDR | 85.4%, 83.9% - 86.9% | 89.4%, 87.1% - 91.7% |
| | PDR | 82.5%, 78.5% - 86.5% | 93.7%, 91.7% - 95.7% |
| 4-class | no DR | 86.3%, 83.9% - 88.7% | 87.8%, 85.9% - 89.7% |
| | mild and moderate NPDR | 81.3%, 77.2% - 85.4% | 84.6%, 82.6% - 86.6% |
| | severe NPDR | 12.0%, 2.0% - 22.0% | 100.0%, 100.0% - 100.0% |
| | PDR | 87.8%, 85.6% - 90.0% | 87.1%, 85.1% - 89.1% |

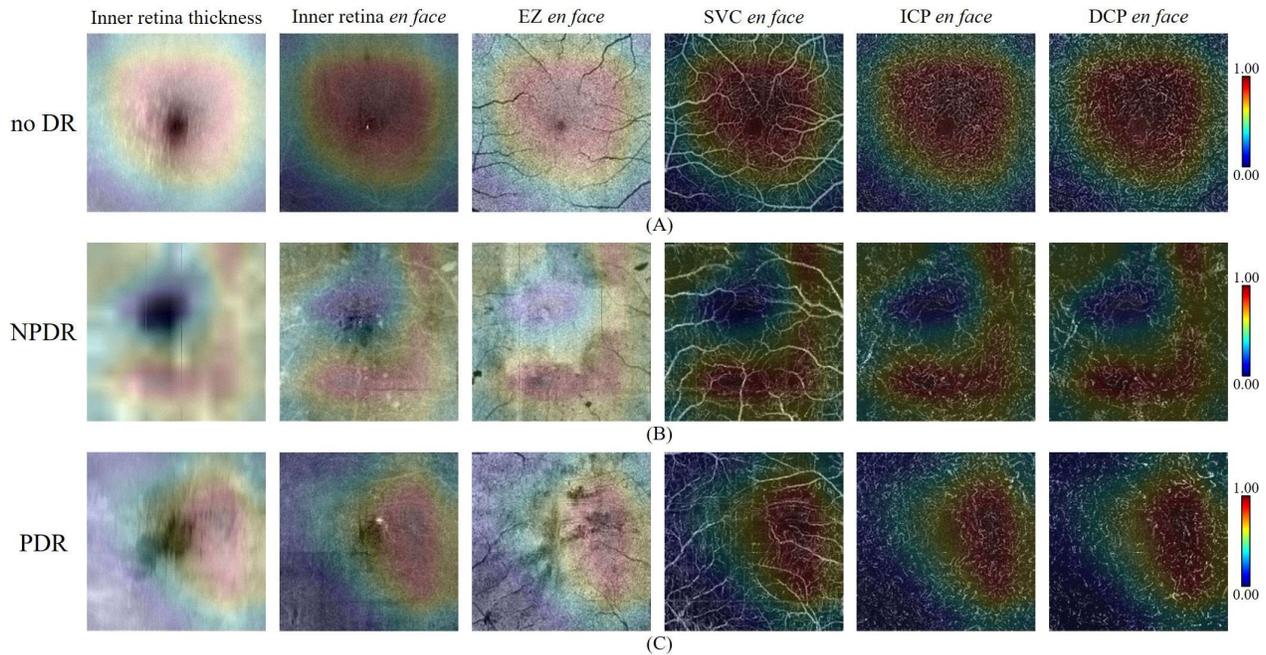

Fig. 5. The CAMs of three correctly predicted cases with different DR classes. In each row, the inner retina thickness map, inner retinal *en face* OCT, EZ *en face* OCT, SVC *en face* OCTA, ICP *en face* OCTA, and DCP *en face* OCTA were overlaid by the corresponding CAMs. In addition, the color bar of each CAM was on the right side of each row. (a) CAMs of case without DR. (b) CAMs of a case with NPDR. (c) CAMs of a case with PDR.

To further quantitatively analyze the proposed method, we performed five comparisons on our local dataset to investigate the accuracy and stability of the proposed DR classification framework. First, we compared the performance of the network trained on combined OCTA and OCT structural data inputs to the network trained on either structural OCT or OCTA data separately. Second, we compared the performances of our network with no dropout, standard dropout (0.2 dropout rate), and proposed adaptive dropout. Third, we compared the performances of our network with traditional class weight balancing and proposed adaptive label smoothing. Fourth, we compared the performances of different network architectures (*ResNet* [18], *DenseNet* [29], *EfficientNet* [36], VGG16 [37], VGG19 [37], ResNet-v2 [38], Inception-v4 [39] and the proposed *DcardNet*) with or without the adaptive label smoothing. Finally, we compared the performances of our method with a previously proposed ensemble network [17] on the 2-class DR classification. In addition, all the results (including ours) in the comparisons below (sections *A, B, C, D* and *E*) were based on 5-fold cross-validation with 20% exclusively reserved for testing.

### A. Comparison between the three input patterns

The inputs had six channels obtained from both OCT and OCTA data. In order to verify the necessity of this input design, comparison of classification accuracies between the OCT-based inputs, OCTA-based inputs, and OCT+OCTA-based inputs were performed. The network used a set of 6 *enface* images as input. From structural OCT-based we included an inner retina thickness map, an inner retina average projection, and an EZ average projection. The OCTA-based inputs are *enface* maximum projection of the SVC, ICP, and DCP. Table V shows the overall accuracies of the three levels of DR classification based on three different input patterns. Compared to the OCT-based input, the proposed input design greatly increased ($\approx$ 10%) the overall accuracies of 3 and 4-class DR classification. Compared to the OCTA-based input, the overall accuracies also increased for 3-class DR classification. For the 4-class DR classification, though the overall accuracy of OCT+OCTA-based was the same as only OCTA-based, the sensitivities of OCT+OCTA-based shown in Table VI were more balanced than only OCTA-based. For the 2-class DR classification, which has the same accuracy based on three different input patterns, the CAMs only based on OCT and OCTA were both calculated to study the different influences from OCT and OCTA (Fig. 6). Through first row, we can see the CAMs only based on OCT were both convex polygons centered on the fovea of nrDR and rDR eyes. On the contrary, the two CAMs only based on OCTA were quite different and have more complicated shapes. This comparison shows that more detailed information was used in the DR classification only based on OCTA.

TABLE V
COMPARISON OF THE DR CLASSIFICATION ACCURACIES AT MULTIPLE LEVELS BETWEEN THREE DIFFERENT INPUT PATTERNS

| Inputs patterns | 2-class (mean, 95% CI) | 3-class (mean, 95% CI) | 4-class (mean, 95% CI) |
|---|---|---|---|
| OCT-based | 94.2%, 91.1% - 97.3% | 63.7%, 60.4% - 67.0% | 54.7%, 52.1% - 57.3% |
| OCTA-based | 94.2%, 90.5% - 97.9% | 74.0%, 69.7% - 78.3% | 64.7%, 61.5% - 67.9% |
| OCT+OCTA-based | 94.2%, 91.9% - 96.5% | 76.7%, 73.4% - 80.0% | 64.7%, 61.5% - 67.9% |

TABLE VI
COMPARISON OF THE SENSITIVITIES AND SPECIFICITIES OF FOUR DR SEVERITIES BETWEEN THREE DIFFERENT INPUTS PATTERNS

| DR severities | | OCT-based (mean, 95% CI) | OCTA-based (mean, 95% CI) | OCT+OCTA-based (mean, 95% CI) |
|---|---|---|---|---|
| no DR | Sensitivity | 80.0%, 75.4% - 84.6% | 84.7%, 80.1% - 89.3% | 82.4%, 77.3% - 87.5% |
| | Specificity | 77.2%, 75.5% - 78.9% | 84.2%, 82.5% - 85.9% | 85.1%, 82.8% - 87.4% |
| mild and moderate NPDR | Sensitivity | 36.3%, 31.7% - 40.9% | 63.8%, 59.2% - 68.4% | 66.2%, 63.2% - 69.2% |
| | Specificity | 80.5%, 78.2% - 82.8% | 82.3%, 79.7% - 84.9% | 81.8%, 78.6% - 85.0% |
| severe NPDR | Sensitivity | 0.0%, 0.0% -0.0% | 2.0%, 0.0% - 5.9% | 4.0%, 0.0% - 8.8% |
| | Specificity | 100.0%, 100.0% - 100.0% | 100.0%, 100.0% - 100.0% | 100.0%, 100.0% - 100.0% |
| PDR | Sensitivity | 78.8%, 76.0% - 81.6% | 82.4%, 77.3% - 87.5% | 81.2%, 76.9% - 85.5% |
| | Specificity | 81.4%, 80.0% - 82.8% | 85.1%, 82.8% - 87.4% | 85.6%, 83.9% - 87.3% |

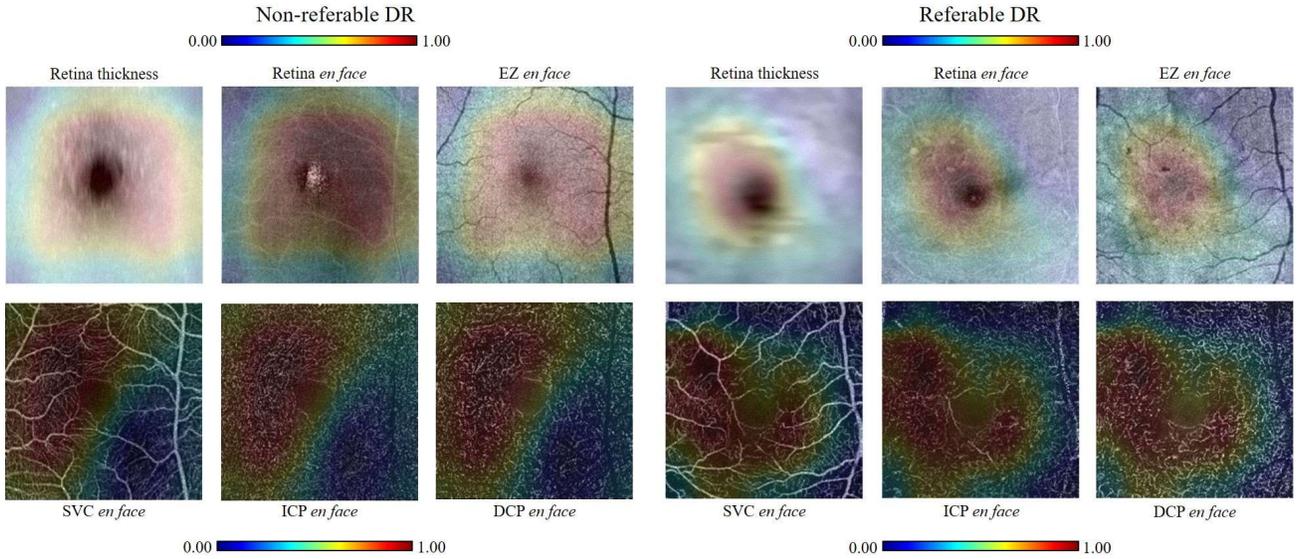

Fig. 6. Comparison between CAMs generated from the two-class DR classification only based on OCT or OCTA. First row: CAMs from the OCT-only network overlaid on the three *en face* OCT layers scanned from nrDR and rDR eyes. Second row: CAMs from the OCTA-only network overlaid on the corresponding OCTA.

Table VI summarizes the comparison of the sensitivities and specificities between the three input patterns and 4 different DR classes. The combined input design improved the sensitivities of two intermediate severity classes. While the overall accuracies of OCTA-based input and OCT+OCTA-based input were the same, using OCT+OCTA based input reduced the variation of sensitivities between different DR severities.

### B. Comparison between different dropout strategies

The performances comparison between our network with three different dropout strategies were shown in Table VII. Proposed network with adaptive dropout shown the highest accuracies in all three DR classification levels. The accuracy increasing based on adaptive dropout was most obvious in the 3-class DR classification.

TABLE VII
COMPARISON OF THE OVERALL ACCURACY BETWEEN THREE DIFFERENT DROPOUT STRATEGIES

| Dropout strategies | 2-class (mean, 95% CI) | 3-class (mean, 95% CI) | 4-class (mean, 95% CI) |
|---|---|---|---|
| no dropout | 93.6%, 91.7% - 95.5% | 73.3%, 71.9% - 74.7% | 64.3%, 62.6% - 66.0% |
| Standard dropout (0.2) | 94.2%, 90.5% - 97.9% | 75.3%, 73.4% - 77.2% | 64.3%, 62.6% - 66.0% |
| Adaptive dropout | 94.2%, 91.9% - 96.5% | 76.7%, 73.4% - 80.0% | 64.7%, 61.5% - 67.9% |

### C. Comparison between class weight balancing and adaptive label smoothing

To gauge the ability of adaptive label smoothing to compensate for the unbalanced classes in our data set, we compared the performance of our network with class weight balancing, adaptive label smoothing, or both (Table VIII). At each classification level, the network trained with adaptive label smoothing outperformed both class weight balancing and the network using both class weight and adaptive label smoothing.

TABLE VIII
COMPARISON OF THE OVERALL ACCURACY BETWEEN THREE DIFFERENT WEIGHT BALANCING STRATEGIES

| Weight balancing strategies | 2-class (mean, 95% CI) | 3-class (mean, 95% CI) | 4-class (mean, 95% CI) |
|---|---|---|---|
| Class weight balancing | 93.6%, 91.7% - 95.5% | 75.3%, 72.7% - 77.9% | 64.3%, 61.7% - 66.9% |
| Adaptive label smoothing | 94.2%, 91.9% - 96.5% | 76.7%, 73.4% - 80.0% | 64.7%, 61.5% - 67.9% |
| Both strategies | 94.2%, 1.9% - 96.5% | 76.0%, 74.2% - 77.8% | 63.9%, 61.3% - 66.5% |

TABLE IX
COMPARISON OF THE OVERALL ACCURACIES BETWEEN DIFFERENT ARCHITECTURES WITH OR WITHOUT ADAPTIVE LABEL SMOOTHING

| Architectures | Label pattern | 2-class (mean, 95% CI) | 3-class (mean, 95% CI) | 4-class (mean, 95% CI) |
|---|---|---|---|---|
| *ResNet*-18 [18] | Normal label | 92.9%, 91.7% - 94.1% | 71.7%, 69.9% - 73.5% | 64.0%, 61.3% - 66.7% |
| | Adaptive label | 93.6%, 90.9% - 96.3% | 75.3%, 73.4% - 77.2% | 64.3%, 62.1% - 66.5% |
| *DenseNet*-53 [29] | Normal label | 91.5%, 90.4% - 92.6% | 72.0%, 70.8% - 73.2% | 64.3%, 62.1% - 66.5% |
| | Adaptive label | 91.9%, 90.7% - 93.1% | 73.3%, 72.3% - 74.3% | 64.3%, 62.6% - 66.0% |
| *EfficientNet*-B0 [36] | Normal label | 91.9%, 90.0% - 93.8% | 70.3%, 68.4% - 72.2% | 60.7%, 59.0% - 62.4% |
| | Adaptive label | 92.9%, 91.7% - 94.1% | 73.7%, 72.5% - 74.9% | 61.7%, 60.3% - 63.1% |
| *VGG16* [37] | Normal label | 87.1%, 86.7% - 88.9% | 71.0%, 68.3% - 73.7% | 64.4%, 62.4% - 66.2% |
| | Adaptive label | 89.5%, 86.1% - 92.9% | 71.7%, 67.9% - 75.5% | 66.2%, 61.4% - 71.1% |
| *VGG19* [37] | Normal label | 89.8%, 88.2% - 91.5% | 72.7%, 67.8% - 77.5% | 61.6%, 59.0% - 64.3% |
| | Adaptive label | 90.8%, 87.5% - 94.2% | 74.7%, 69.3% - 80.0% | 63.9%, 59.4% - 68.5% |
| *ResNet-v2*-50 [38] | Normal label | 89.8%, 88.5% - 91.2% | 74.0%, 71.6% - 76.4% | 64.6%, 62.4% - 66.7% |
| | Adaptive label | 90.5%, 89.0% - 92.0% | 76.0%, 73.5% - 78.5% | 65.9%, 63.0% - 68.8% |
| *Inception-v4* [39] | Normal label | 89.2%, 86.6% - 91.7% | 68.7%, 64.3% - 73.0% | 57.7%, 54.9% - 60.5% |
| | Adaptive label | 90.2%, 86.5% - 93.9% | 72.7%, 69.0% - 76.3% | 62.0%, 60.1% - 63.9% |
| *DcardNet*-36 | Normal label | 93.6%, 91.7% - 95.5% | 74.7%, 73.5% - 75.9% | 64.3%, 62.6% - 66.0% |
| | Adaptive label | 94.2%, 91.9% - 96.5% | 76.7%, 73.4% - 80.0% | 64.7%, 61.5% - 67.9% |

*D. Comparison between different network architectures*

We also compared the performances of *ResNet*-18, *EfficientNet*-B0, and *DenseNet*-53, *VGG16*, *VGG19*, *ResNet-v2*-50, *Inception-v4* and proposed *DcardNet*-36 with or without adaptive label smoothing for the DR classification at multiple levels on the same dataset. Among them, *DenseNet*-53 is a modified *DenseNet* architecture with 53 layers (52 convolution and 1 dense layers) which achieved the highest accuracy compared to other *DenseNet* architectures. In addition, no transfer learning was used in the training of all the networks above and all the final models were trained from scratch with empirically selected optimal hyper-parameters.

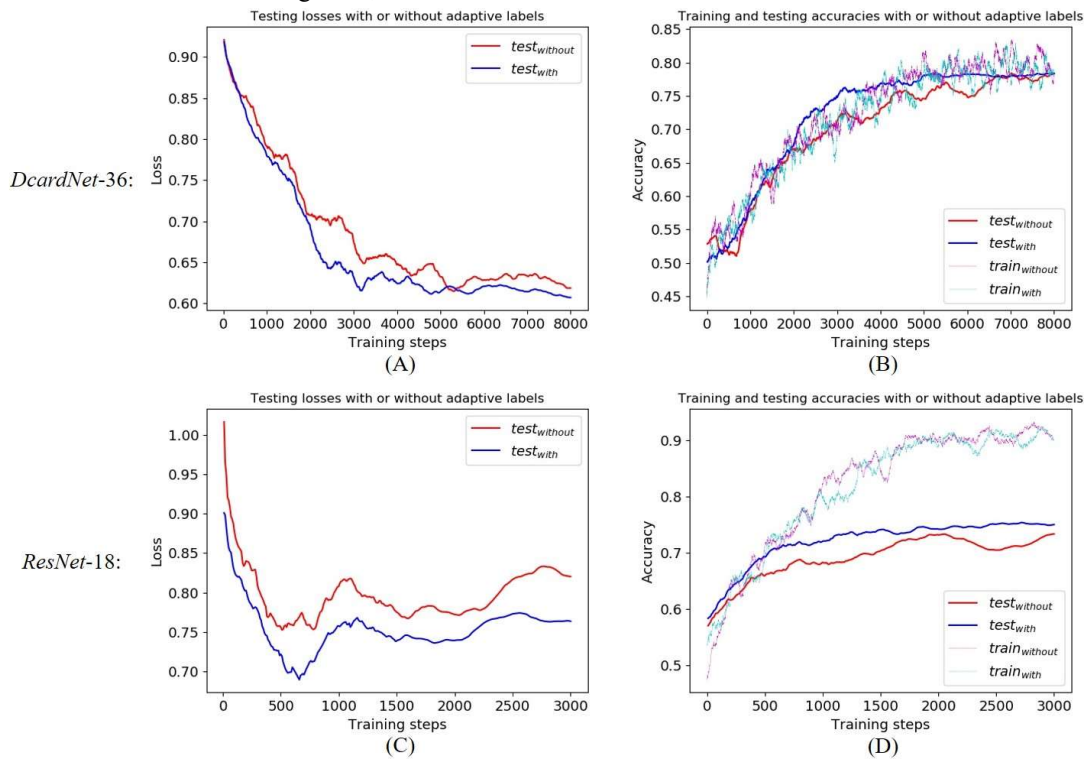

Fig. 7. Comparisons of the losses and accuracies based on proposed *DcardNet*-36 and *ResNet*-18 with or without adaptive label smoothing on the 3-class dataset with 20% of the data as the testing dataset. (A) Comparisons of the testing losses based on *DcardNet*-36. (B) Comparisons of the training (dotted lines) and testing (solid lines) accuracies based on *DcardNet*-36. (C) Comparisons of the testing losses based on Res*Net*-18. (D) Comparisons of the training (dotted lines) and testing (solid lines) accuracies based on *ResNet*-18.

Table IX shows the overall accuracies of the three levels of DR classification based on all eight network architectures. Our network architecture with or without adaptive label smoothing achieved the highest accuracies on both 2-class and 3-class DR classifications. Only the 4-class DR classification accuracies of *VGG16* and *ResNet-v2*-50 were about 1% higher than ours. In addition, the use of the proposed adaptive label smoothing improved the classification accuracies of all architectures.

To further analyze the improvement in generalization by the adaptive label smoothing, we measured the losses and accuracies based on the proposed *DcardNet*-36 and *ResNet*-18 with or without adaptive label smoothing on the 3-class dataset with 20% data exclusively used as testing dataset (Fig. 7). The testing losses and accuracies were obtained after each 10 training steps and both smoothed by an average filter with length 50. The training accuracies were smoothed by an average filter with length 100. In Fig. 7A and 7C, we can see the testing losses with adaptive label smoothing were lower than the losses without adaptive label smoothing during the entire training process. Though the training accuracies with and without adaptive label smoothing were almost the same, the testing accuracies with adaptive label smoothing were always higher than the accuracies without adaptive label smoothing (Fig. 7B and 7D). In addition, the testing accuracy with adaptive label smoothing increased more smoothly and monotonically than the accuracy without adaptive label smoothing. By comparing two rows, we can also intuitively see that *DcardNet*-36 has better generalization performance and lower overfitting than the *ResNet*-18. And as noted, the adaptive label smoothing has higher improvement on *ResNet*-18 than *DcardNet*-36.

### E. Comparison with ensemble networks based on enface OCT and OCTA

We also compared the performances on 2-class DR classification between our method and a previously proposed ensemble network [17] which also uses *enface* OCT and OCTA as inputs. The ensemble network consisted of four VGG19 [37] with pre-trained ImageNet parameters. The inputs of the ensemble network were SVC and DCP *enface* images respectively generated from OCT and OCTA. Based on the same implementation details, the results of the ensemble network were shown in Table X. The overall accuracy, sensitivities and specificities of our method are all better than the ensemble network.

TABLE X
COMPARISON OF THE 2-CLASS DR CLASSIFICATION PERFORMANCE BETWEEN OUR METHOD AND THE ENSEMBLE NETWORK

| Methods | Accuracy (mean, 95% CI) | Sensitivity of rDR (mean, 95% CI) | Specificity of rDR (mean, 95% CI) |
|---|---|---|---|
| Ensemble network | 86.8%, 85.3% - 88.2% | 90.5%, 84.8% - 92.6% | 78.9%, 73.1% - 88.4% |
| Our method | 94.2%, 91.9% - 96.5% | 96.0%, 94.2% - 97.8% | 90.5%, 87.1% - 94.0% |

## VI. DISCUSSION

We proposed a new convolutional neural network architecture based on dense and continuous connection with adaptive rate dropout (*DcardNet*) for automated DR classification based on OCT and OCTA data. To our knowledge this is the first study to report DR classification across multiple levels based on OCT and OCTA data. A classification scheme like this is desirable for several reasons. OCT and OCTA are already an extremely common procedures in ophthalmology [40]. An automated DR classification framework could further extend the applications of these technologies. If OCT/OCTA can deliver the same diagnostic value as other modalities, the number of procedures an individual would require for accurate diagnosis would be reduced, which will ultimately lower clinical burden and healthcare costs. Furthermore, OCT/OCTA provide a unique set of features (three-dimensionality combined with high-resolutions) that may prove to have complimentary or superior diagnostic value for some diseases; however, the sheer size of OCT/OCTA data sets inhibits detailed analysis. By providing tools for automation, we can begin to acquire data that can help identify new biomarkers or other features useful for DR staging.

Our network design incorporated several ideas that enabled rapid training and accurate results. We found that, compared to the residual structure, the dense connected structure was much more resistant to overfitting. However, the dense connection also had a lower convergence rate than the residual structure (*ResNet*). In order to increase the convergence rate and keep overfitting low, the dense and continuous connection was proposed and used in this study. In the new architecture, a dense connection was continuously used within a sliding window from the first bottleneck block to the last one. Compared to use of dense connections within each block (*DenseNet*), the new structure was able to deliver useful features with lower losses. In addition, the use of dropout with adaptive rate kept overfitting low. Sixteen bottleneck blocks with 24 output features were finally chosen in this study based on the classification complexity and size of the dataset. For more classes and larger datasets (like those seen in ImageNet), more bottleneck blocks with more output features may be needed.

Adaptive label smoothing was proposed and used to reduce overfitting in this study. The labels of each of the training steps were adaptively smoothed based on their prediction histories. Because of the adaptively smoothed labels, the convergence of the network could be more focused on the mispredicted data, rather than the data that was already correctly predicted. The only concern for this technique is the inaccuracy introduced from data which have an ambiguous ground truth. Therefore, this technique is more suitable to well-labeled datasets. Another technique we used to reduce the overfitting was data augmentation, which has been widely used in medical image classification. In addition to improving data diversity, the data augmentation we used in this study also fits with practical diagnosis, where the doctors' diagnosis is not influenced by the angle of the *en face* vasculature.

For practical and historical reasons, layer segmentation has become a necessary step for most analytic pipelines using OCT and OCTA. The enface images based on segmented layers are not only used to automated DR classification but also necessary for OCT-based routine diagnosis. From a machine learning perspective, this is a mixed blessing. Dimensionality reduction enables swifter training (since 3D data sets are much sparser), but simultaneously suppresses otherwise learnable information. Our network was trained on datasets segmented using manually corrected software [23, 41-44], which introduces both a manual step into our data pipeline and some idiosyncrasy into ground truth. State of the art layer segmentation now requires less manual correction [45-48], and we believe will continue to do so. However, the accuracy of our results is, unfortunately, probably negatively impacted by these limitations in the ground truth used for training. OCTA networks are also unfortunately limited by a relative paucity of data compared to other medical imaging datasets. As more OCTA data is acquired, training on 3D data volumes may become practicable, mitigating this concern.

The overall accuracies based on OCT-based inputs, OCTA-based inputs, and OCT+OCTA-based were the same of 2-class and 4-class DR classification. However, we still think the OCT+OCTA-based input is a better option. First, this input strategy still improved the overall accuracy of 3-class DR classification and also balanced the sensitivities of 4-class DR classification. Second, some DR or DME related biomarkers such as fluid could be easier detected in OCT. At last, the OCT *enface* generation is not time-consuming after the retinal layers are segmented, and this segmentation is also needed for OCTA *enface* generation. Therefore, the designed OCT+OCTA-based input pattern is still preferable for the DR classification.

The overall accuracy of the 4-class DR classification was much lower than other two classification levels. In addition, the sensitivity of severe NPDR classification was much lower than the other classes. These two issues are caused by the small differences between the two NPDR classes, which are much smaller than the differences between no DR, NPDR and PDR. Another reason for this relatively low performance is that the number of severe NPDR cases was much smaller than other classes. Therefore, the network could hardly identify the differences between two NPDR severities before overfitting sets in. In future work, we will focus on overcoming these problems by using a larger and more balanced dataset and adding some extra manually selected biomarkers to the inputs. In addition, according to the difference between accuracies based on 5-fold and 10-fold cross-validations, using "leave-one-subject-out" experiments could also help increase the final accuracy and sensitivity.

Compared to CFP-based DR classifications [9-12], the overall accuracy of our 2-class DR classification was slightly lower. One reason was that the CFP-based DR classifications had about 100 times as much data as we did. Though we have stratified accuracies on 2-class and 3-class DR classifications based on our relatively small dataset, a huge dataset like those available from CFP could further improve our DR classification to state-of-art performance. Furthermore, the current classification used for training our algorithm, which is based on grading from color fundus photography, may not be optimal for OCTA classification. The current gold standard for DR diagnosis is based on color fundus photograph which is a considerably different modality from OCT/OCTA. Features used to distinguish some DR classifications using the ETDRS scheme may be missing from OCT/OCTA datasets, which could hurt the accuracy of our algorithm.

Furthermore, there are currently trade-offs between CFP and OCTA. CFP provides a larger field of view, but at lower resolution and the cost of a dimension of information when compared to OCTA. Both provide visualization of a unique set of pathological features. Currently, CFP can provide some information that is inaccessible to OCTA, though complimentary features of the same pathology may be visible to OCTA [49, 50]. However, we do not conceive of this work solely as a means to automatize through OCTA grading what can already also be automated through CFP. Instead, we believe that this work demonstrates that the feature set that can be extracted through OCTA images of the macular region is sufficient to diagnose DR at a level similar to CFP, without relying on the specific features (microaneurysms, bleeding) provided by CFP. We think this this is innovative of its own accord because it adds value to an existing technology.

We note additionally that the amount of data procured from structural OCT in conjunction with OCTA is much larger than that from CFP, by virtue of being high-resolution and three-dimensional. Features like microaneurysms that are currently used to stage DR may not end up being essential to DR staging, as our work shows. Close parity with ETDRS grading of CFP data indicates significant potential for OCTA staging as OCTA hardware continues to improve.

## VII. CONCLUSION

In conclusion, we proposed a densely and continuously connected convolutional neural network with adaptive rate dropout to perform a DR classification based on OCT and OCTA data. Among our architecture designs, the dense and continuous connections improved the convergence speed and adaptive rate dropout reduced overfitting. Three classification levels were finally performed to fulfill requests from clinical diagnosis. In addition, adaptive label smoothing was proposed and used in this study. With the addition of adaptive label smoothing, the convergence of the network could be more focused on the mispredicted data, rather than the data that was already be correctly predicted. In the end, the trained model focused more on the common features of the whole dataset, which also reduced overfitting. Classifying DR at three levels and generating CAMs could both help clinicians improve diagnosis and treatment.